\documentclass[aps,fleqn,superscriptaddress]{revtex4}
\usepackage{graphicx,color,natbib}
\usepackage{amsmath,amssymb,amsfonts}

\newcommand{\bse}{\begin{subequations}}
\newcommand{\ese}{\end{subequations}}
\newcommand{\be}{\begin{equation}}
\newcommand{\ee}{\end{equation}}
\newcommand{\bea}{\begin{eqnarray}}
\newcommand{\eea}{\end{eqnarray}}
\newcommand{\ba}{\begin{array}}
\newcommand{\ea}{\end{array}}
\newcommand{\h}{\frac{1}{2}}

\def\B{{\mathcal{B}}}

\def\m{\frac{\mu}{T}}
\input amssym.def
\input amssym.tex

\usepackage[colorlinks=true, linkcolor=blue, bookmarks=true]{hyperref}

\begin{document}
IPM/P-2018/078 

\title{Evolution of Holographic Complexity Near Critical Point}

\author{H. Ebrahim\footnote{hebrahim@ut.ac.ir}}
\affiliation{Department of Physics, University of Tehran, North Karegar Ave., Tehran 14395-547, Iran}
\affiliation{School of Physics, Institute for Research in Fundamental Sciences (IPM),
P.O.Box 19395-5531, Tehran, Iran}
\author{M. Asadi\footnote{$\rm{m}_{-}$asadi@ipm.ir}}
\affiliation{School of Physics, Institute for Research in Fundamental Sciences (IPM),
P.O.Box 19395-5531, Tehran, Iran}
\author{M. Ali-Akbari\footnote{$\rm{m}_{-}$aliakbari@sbu.ac.ir}}
\affiliation{Department of Physics, Shahid Beheshti University G.C., Evin, Tehran 19839, Iran}

\begin{abstract}
The holographic complexity has been studied in a background which includes a critical point in the dual field theory. We have examined how the complexity rate and the saturation time of dynamical variables in the theory behave as one moves towards the critical point. Two significant results of our analysis are that (i) it takes more time for the complexity in field theory dual to become time dependent as one moves away from the critical point and (ii) near the critical point the complexity starts evolving linearly in time sooner than the other points away from it. We also observe different behaviour for complexity rate in action and volume prescriptions. In action prescription we have used the time scales in theory to obtain the dynamical critical exponent and interestingly have observed that different time scales produce the same value up to very small error. 
\end{abstract}

\maketitle

\tableofcontents

\section{Introduction and Results}

Recently new connections between quantum information and quantum gravity have been developed which has attracted a lot of attention in the literature. A promising framework to investigate these connections is AdS/CFT correspondence, holography,  or more generally gauge/gravity duality \cite{Maldacena:1997re}. This framework has been applied to study quantities such as entanglement entropy and mutual information and recently extended to the quantum computational complexity in field theory. The quantum complexity is the minimum number of elementary operations (quantum gates) needed to produce an arbitrary state which is of interest from a fixed reference state, for example see \cite{Aaronson:2016vto} for the review.  

Quantum complexity in AdS/CFT picture sheds light on physics behind the horizon. In fact it was proposed that the quantum information complexity of the boundary field theory state is encoded geometrically in the gravitational dual background, or more specifically the interior geometry of the black hole \cite{Susskind:2014rva}. A primary example is the eternal two-sided AdS-Schwarzchild black hole which is dual to the thermofield double state in the boundary theory \cite{Maldacena:2001kr}. This is an entangled state of two copies of the boundary conformal field theory. This entanglement is responsible for the geometric connection in the bulk \cite{Maldacena:2013xja}. More specifically it has been shown that the black hole interior grows in time long after the equilibrium is reached \cite{Hartman:2013qma}. On the other hand one expects the complexity of the dual thermal plasma to increase long after local equilibrium is reached. In fact the complexity of a state can increase due to its Hamiltonian time evolution and this growth is dual to the late-time growth of the interior. It is conjectured that the complexity continues to grow up to a time scale exponential in the number of degrees of freedom \cite{Brown:2017jil}.  Therefore it is proposed that these two descriptions belong to the same phenomena \cite{Susskind:2014rva}.

This proposal has been introduced in two different recipes: to characterize the size of black hole interior with its spatial volume or its action are known as volume \cite{Stanford:2014jda}  or action \cite{Brown:2015bva,Brown:2015lvg} conjectures, CV or CA, in the literature. The volume complexity prescription states that
\be
{\cal{C}}_V=\frac{V}{G_N l} ,
\ee
where $V$ is the volume of the maximal codimension-one bulk time slice anchored at boundaries at some specific times and $l$ is a length scale associated with the geometry. $G_N$ is the Newton's constant.  One of the drawbacks of volume prescription is the extra free length scale, $l$, which needs to be fixed by hand. Once the length scale is fixed by considering any particular black hole the predictions can be made for the other cases. The complexity equals action includes almost all good features of volume conjecture and in addition one does not need to fix an extra free length scale. The action complexity prescription equates the boundary theory complexity with the gravitational action evaluated on a space-time region called Wheeler-de Witt (WdW) patch: 
\be
{\cal{C}}_A=\frac{I_{WdW}}{\pi \hbar} .
\ee
This region is bounded by null surfaces that reach the relevant times on the left and right boundaries. Since holographic complexity has been proposed, a lot of papers have studied different aspects of it \cite{Alishahiha:2018lfv}.  

An interesting question that can be asked is about the late-time growth rate of complexity. This proposal asserts that it is $\frac{2 \hat{M}}{\pi}$ where $\hat{M}$ is proportional to the mass of the black hole. Therefore it is independent of space-time dimension or other information of the boundary theory \cite{Brown:2015bva,Brown:2015lvg}. 

In this paper we would like to study the full time evolution of holographic complexity of the charged thermofield double state and its behaviour near the critical point. Therefore we consider a field theory with a critical point whose gravity dual is a charged black hole, discussed in the upcoming section. The field theory is characterized by temperature $T$ and chemical potential $\mu$ and its phase diagram contains a first order phase transition line which ends in a critical point. It is characterized by the ratio of $\frac{\mu}{T}$, as expected, since the underlying theory is conformal. Although there are known backgrounds in holographic framework with a critical point in the field theory dual, we choose this background since it can be investigated analytically and one has more control on the calculations \cite{DeWolfe:2010he}. 

We try to check the possible quantities that are interesting to analyze in this background near the critical point. The results of our studies can be summarized as follows: 
\begin{itemize}
\item In action prescription for complexity, CA, the behaviour of the critical time, the time at which the complexity rate becomes non-zero, with respect to $\m$ is studied. The critical time decreases as $\m$ increases and gets its minimum value at the critical point. It means that it takes more time for the complexity in field theory dual to become time dependent as one moves away from the critical point. 
\item The above behaviour persists for the other time scales in theory such as the saturation time of $r_m$, $t_{rs}$, and the saturation time of complexity rate in action prescription, $t_{cs}$. $r_m$ is the value of $r$ at which the past null rays of WdW patch intersect. Therefore near the critical point the complexity starts evolving in time linearly sooner than the other values of $\m$. 

In volume prescription, CV, the saturation time of complexity rate does not vary much with $\m$ and is almost constant. 
\item In CA we also obtain the dynamical critical exponent using the aforementioned time scales and they all are equal up to less than one precent error. 
\begin{table}[ht]
\caption{value of dynamical critical exponent obtained from different time scales}
\vspace{1 mm}
\centering
\begin{tabular}{c c c c}
\hline\hline
~~$ \rm{time\  scale} $ ~~   &~~$ \theta $ ~~   &   ~~ $ \rm{relative\ error} $ ~~ & ~~ $ \rm{rms} $~~ \\[0.5ex]
\hline
$t_c$ & -0.44 & 12 \% & 0.07 \\
$t_{rs}$ & -0.46 &  8 \%  & 0.22 \\
$t_{rc}$ & -0.48 &  4 \% & 0.02 \\
\hline
\end{tabular}\\[1ex]
\label{list}
\end{table}
The relative error is defined as $\frac{|\theta-0.5|}{0.5}$ where 0.5 is the value of dynamical critical exponent obtained from conserved current and quasinormel modes \cite{DeWolfe:2011ts,Finazzo:2016psx}. The last row in table \ref{list} is the best match with the known result 0.5.  

Regarding the result for CV mentioned in the previous item we can not obtain dynamical critical exponent in CV prescription. 
\item We have also examined the difference between $t_{rs}$ and $t_{cs}$ in CA prescription and conclude that this difference reaches zero as one moves towards the critical point. This means that the complexity rate and $r_m$ saturate at the same time at the critical point. For the other values of $\m$ the complexity rate saturates later than $r_m$. Although different terms may contribute in time dependence of complexity rate, it seems that at the critical point the contribution of time dependence of $r_m$ is dominant. 
\item We also show that the late-time values of $r_m$ monotonically decreases moving towards the critical point in $\m$ parameter space while complexity rate in action prescription increases at first and as getting closer to the critical point peaks and starts decreasing. In contrast the late time values of complexity rate in volume prescription monotonically decreases moving towards the critical point. 
\end{itemize}

\section{The Charged Black Hole Background with Critical Point}
In this section we review the 5-dim background obtained as a solution to the following gravitational action 
\be
\label{1R action}
S=\frac{1}{16 \pi G_5} \int d^5x \sqrt{-g} \bigg{[}R-\frac{f(\phi)}{4} F_{\mu\nu} F^{\mu\nu}-\h\partial_\mu \phi \partial^\mu \phi-V(\phi) \bigg{]} ,
\ee
where $G_5$ is the five dimensional Newton's constant. This action is the gravitational action of 1RCBH model \cite{Gubser:1998jb} with dilaton potential as 
\be
V(\phi)=-\big{(}8 e^{\frac{\phi}{\sqrt{6}}} + 4 e^{-\sqrt{\frac{2}{3}}\phi}\big{)} ,
\ee
and the Maxwell-Dilaton coupling
\be
f(\phi)=e^{-2 \sqrt{\frac{2}{3}}\phi} .
\ee
The solution to the equations of motion derived from the action \eqref{1R action} can be summarized as
\bea
\label{metric}
ds^2 &=& e^{2 A(r)} (-h(r) dt^2+d{\vec{x}}^2) + \frac{e^{2 B(r)}}{h(r)} dr^2,\\ 
A(r) &=& \text{ln} r + \frac{1}{6} \text{ln} (1+\frac{Q^2}{r^2}) ,\cr
B(r) &=& - \text{ln} r - \frac{1}{3} \text{ln} (1+\frac{Q^2}{r^2}) ,\cr 
h(r) &=& 1-\frac{M^2}{r^2(r^2+Q^2)} ,\cr 
\phi(r) &=& - \sqrt{\frac{2}{3}} \text{ln} (1+\frac{Q^2}{r^2}), \\ 
A_t(r) &=& \bigg{(} - \frac{M Q}{r^2+Q^2}+\frac{M Q}{r_H^2+Q^2}\bigg{)}, 
\eea
where $A_t$ is the time component of the gauge field which has been chosen to be zero at the horizon and regular on the boundary. $M$ is the black hole mass and $Q$ is its charge. We set AdS radius equal to one throughout the paper. The boundary of this asymptotically AdS solution is located at $r \rightarrow \infty$ and $r_H$ is the black hole horizon and is obtained from $h(r_H)=0$,
\be
\label{horizon}
r_H=\sqrt{\frac{\sqrt{Q^4+4 M^2}-Q^2}{2}} .
\ee 
The chemical potential and the temperature in the dual field theory are 
\bea
\label{mu}
\mu&=&\lim_{r \to \infty} A_t = \frac{Q r_H}{\sqrt{Q^2 + r_H^2}} ,\\
\label{temp}
T&=&\frac{\sqrt{-{g_{tt}}' {g^{rr}}'}}{4 \pi}\bigg{|}_{r=r_H} = \frac{Q^2+2 r_H^2}{2 \pi \sqrt{Q^2 + r_H^2}} .
\eea
Having obtained $\frac{Q}{r_H}$ in the bulk solution with respect to $\frac{\mu}{T}$ in the boundary we will see that the background considered here contains two different branches of variables $\frac{Q}{r_H}$ corresponding to each value of $\frac{\mu}{T}$. It indicates the existence of a first order phase transition in field theory. The relation between parameters in the bulk and the parameters in field theory dual can be evaluated using \eqref{mu} and \eqref{temp}
\be
\frac{Q}{r_H}=\sqrt{2} \bigg{(} \frac{1\pm \sqrt{1-(\frac{\sqrt{2}}{\pi}\frac{\mu}{T})^2}}{\frac{\sqrt{2}}{\pi}\frac{\mu}{T}} \bigg{)} .
\ee
Using the relations between entropy, $s$, and charge density, $\rho$, in terms of the bulk solution parameters $Q$ and $r_H$ one can evaluate the Jacobian $\mathcal{J}=\frac{\partial (s,\rho)}{\partial (T,\mu)}$. If the Jacobian is positive (negative) for the set of parameters in one branch the system is thermodynamically stable (unstable). The two branches intersect at the critical point where $\frac{\mu}{T}^*=\frac{\pi}{\sqrt{2}}$ ($\frac{Q}{r_H}=\sqrt{2}$) and the branch with the parameters satisfying $\frac{Q}{r_H}<\sqrt{2}$ is stable. 

Different aspects of this background have been investigated in various papers. For instance in \cite{DeWolfe:2010he} this background has been used as the holographic dual of QCD and the QCD critical point and its exponent has been estimated. Moreover in the dual holographic plasma to this background the bulk viscosity and baryon conductivity have been computed in \cite{DeWolfe:2011ts} and consequently the dynamical critical exponent could be obtained. The value of the dynamical critical exponent was also confirmed in \cite{Finazzo:2016psx} and \cite{Ebrahim:2017gvk} by studying quasi-normal modes and quantum quench in this background, respectively. 

\section{Complexity}\label{csection}
To describe the quantum complexity of states in the boundary field theory two different holographic recipes have been introduced in the literature. One is complexity equals volume conjecture \cite{Stanford:2014jda} and the other one action conjecture \cite{Brown:2015bva,Brown:2015lvg}. In the following we will evaluate both prescriptions for the background we work on in this paper.

\subsection{Complexity=Action Prescription}
\begin{figure}
\includegraphics[width=85mm]{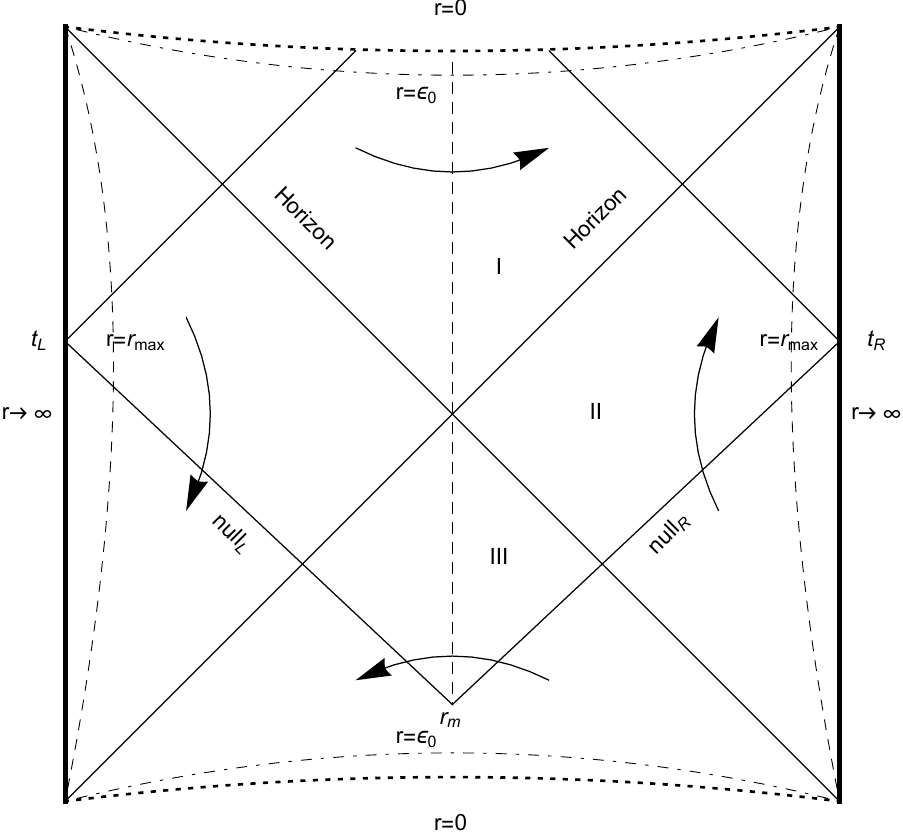}
\caption{WdW patch. The arrows present the time directions on each side.}\label{WdWp}
\end{figure}

Let's consider a quantum state of a particular time slice of the boundary conformal field theory. The action prescription then relates the complexity of this state to the gravitational action evaluated in the corresponding region in the dual gravity theory called "Wheeler-de Witt" (WdW) patch. This region is enclosed by the future and past null sheets sent from the boundary time slice into the bulk space-time. It is illustrated in figure \ref{WdWp} for the asymptotically AdS eternal black hole background where we have called the boundary time on the left and right boundaries as $t_L$ and $t_R$. The boundary state we are considering is on the constant time slices denoted by $t_L$ and $t_R$ on the two asymptotic boundaries, $|\psi(t_L,t_R)\rangle$. This thermofield double state is parametrized by mass $M$, charge $Q$ and chemical potential $\mu$. In fact  for the charged background as well as temperature we have the chemical potential to distinguish the boundary states. Please note that we have set $t_L=t_R\equiv \frac{t}{2}$ in this figure. $t$ stands for the boundary time in which we evaluate the time dependence of the complexity. 

As shown in figure \ref{WdWp}, the time-like killing vectors which correspond to time translation in the background generate upward (downward) flows on the right and left boundaries and hence the action is invariant under shifting the time slices as $t_R+\delta t$ and $t_L-\delta t$. In terms of the boundary theory this corresponds to the invariance of the thermofield double state under an evolution with the Hamiltonian $H_R-H_L$. 

To describe the null sheet boundaries of the WdW patch we work with the tortoise coordinate which is defined as  
\be
\label{rstar}
dr^*=\frac{e^{B(r)-A(r)}}{h(r)} dr ,
\ee
in our background \eqref{metric}. Integrating the above equation we get
\bea
r^*(r) = - \frac{\sqrt{2}}{Q^2} &\bigg{[}& \sqrt{\sqrt{Q^4+4 M^2}-Q^2} ~~{\text{tan}}^{-1} \sqrt{\frac{r^2+Q^2}{\sqrt{Q^4+4 M^2}-Q^2}} \cr
&-&\sqrt{\sqrt{Q^4+4 M^2}+Q^2}~~ {\text{tan}}^{-1} \sqrt{\frac{r^2+Q^2}{\sqrt{Q^4+4 M^2}+Q^2}}~\bigg{]} .
\label{rs}
\eea
Then the Eddington-Finkelstein coordinates, $u$ and $v$ are defined as
\be
v=\tau+r^*(r)~,~~~u=\tau-r^*(r) ,
\ee
where $\tau$ represents the time parametrization of the null sheets which becomes $t$ on the asymptotic boundaries of the background. 

The gravitational action that should be calculated here is

\be\begin{split}\label{actionf}
I&= I_{\text{bulk}} + I_{\text{surface}} + I_{\text{joint}} \cr
&=\frac{1}{16 \pi G_5} \int_{M} d^5x \sqrt{-g} \left(R-\frac{f(\phi)}{4} F_{\mu\nu} F^{\mu\nu}-\h\partial_\mu \phi \partial^\mu \phi-V(\phi)\right)\cr
&+ \frac{1}{8\pi G_5} \int_{\B} d^4x \sqrt{|h|} K + \frac{1}{8\pi G_5} \int_{\Sigma} d^3x \sqrt{\sigma} \eta\cr
&- \frac{1}{8\pi G_5} \int_{\B'} d\lambda d^3\theta \sqrt{\gamma} \kappa + \frac{1}{8\pi G_5} \int_{\Sigma'} d^3x \sqrt{\sigma} a .
\end{split}\ee
The first line is the standard Einstein-Hilbert action in the presence of the cosmological constant $\Lambda=\frac{-6}{L^2}$. The second line contains two terms. The first one is the usual Gibbons-Hawking-York surface term (GHY) for time-like and space-like segments of the boundary which exists on the WdW patch. $K$ in this term is the trace of the extrinsic curvature. The second term gives the contribution from Hayward joint terms which appear in the intersection of the space-like and time-like boundary segments. $\eta$ is defined as the boost angle between the corresponding normal vectors in the intersection. The last line in the action is added due to the presence of null segments in the WdW patch. The first term is the surface term for the null segments of the boundary and the second term is the joint term corresponding to the intersection of the null segments with the other segments. A lot of work has been done on null boundaries in gravity in the literature, such as \cite{Parattu:2015gga}. Please note that we work with the convention in which we choose $\kappa=0$ following \cite{Lehner:2016vdi,Carmi:2017jqz}. Also note that there is no intersection between time-like and space-like surfaces in WdW patch and therefore the Hayward joint term in the second line will be zero. After fixing the convention we are ready to calculate each term individually in our charged black hole background. We will give the details in the appendix \ref{appA} and report here the results. 

It is important to mention that the complexity can be calculated in two regimes which depend on the position of WdW patch. The first regime for which the complexity is proved to be time independent is where the WdW patch is in contact with the past singularity. Due to time independence of complexity in this regime we will not go through the details of the calculation and refer the interested reader to the papers such as \cite{Carmi:2017jqz}, in which this has been done in thorough details. 

The second regime is where the past null sheets from the right and left boundaries intersect before crossing the singularity. As we will show later on, and has been shown in related papers such as \cite{Carmi:2017jqz}, the complexity is time-dependent in this regime. Therefore we have only plotted the WdW patch which results in the time-dependent part of the complexity of the quantum state in the boundary in figure \ref{WdWp}. 

One can easily find the time that separates these two regimes which is called critical time $t_c$,  respecting the convention used in the literature. Using the equation of the past null sheets 
\bea
\label{null sheets}
\text{null}_{\text{L}} :&& \tau_L=-t_L+r^*_\infty -r^*(r) , \cr
\text{null}_{\text{R}} :&& \tau_R=t_R-r^*_\infty +r^*(r) ,
\eea
the critical time is
\be
t_c=2 (r^*_{\infty} - r^*_0)  ,
\label{critical t}
\ee
where $r^*_\infty$ and $r^*_0$ are calculated at the boundary where $r\rightarrow \infty$ and the singularity at $r\rightarrow 0$, respectively. $t_c$ is the boundary time at which the null sheets from left and right halves of the WdW patch intersect at the past singularity; i.e. $(\tau_L=\tau_R)_{r\rightarrow 0}$. $\text{null}_{\text{L}}$ and $\text{null}_{\text{R}}$ lines are shown in figure \ref{WdWp}.

Having obtained $t_c$ we are in the position to calculate the value of $r_m$ where the two past null sheets from the right and left boundaries intersect before hitting the singularity (as shown in figure \ref{WdWp}). Similar to the calculations resulted in \eqref{critical t} using the equations of these null sheets \eqref{null sheets} we obtain
\be
\label{rstar m}
r^*(r_m) = -\frac{t}{2} + r^*_\infty .
\ee
It should be emphasized that $r_m$ changes as the boundary time increases and moves towards the horizon in a sufficiently long time. Therefore it is necessary to see how this meeting point $r_m$ evolves with time. Differentiating \eqref{rstar m} with respect to time we get   
\be\label{eq1}
\frac{d r_m}{dt} = -\h h(r) e^{A(r)-B(r)}\bigg{|}_{r=r_m} .
\ee
This relation suggests that at late times where $r_m$ reaches the horizon, the growth rate of $r_m$ goes to zero as $h(r_H)$ is zero and $e^{A(r)-B(r)}$ reaches a finite value at the horizon. This is confirmed in the next section where we report the numerical results. 

So far we have been trying to obtain some general idea about WdW patch in the background we are interested in. We can start calculating the complexity in the field theory dual to this background using action prescription. As it was mentioned before since we are interested in the time dependence of complexity we only report the calculations and results for the boundary time $t>t_c$. The non-vanishing contributions to the action on WdW patch can be divided into three sets of terms; bulk, GHY and null joint contributions. Due to the symmetry between the left and right halves of the WdW patch, we can calculate the action in the right side of the patch and then multiply it by two. We give the details of the calculations in appendix \ref{appA} and report the time-dependent results here. The complexity in our background becomes
\be
{{\cal{C}}_A}=\frac{I_{WdW}}{\pi \hbar}=\frac{1}{\pi \hbar}(I_{\text{bulk}}+I_{\text{surf}}+I_{\text{joint}}) .
\ee
Therefore the complexity rate is
\be\label{rate}\begin{split}
\frac{d C_A}{dt} &= \frac{16\pi^2 G_5} { V^{(3)}} \frac{d\mathcal{C}_A}{dt}\cr
&=\int_{0}^{r_m} e^{4 A+B}\big(\frac{2}{3}V(\phi)
+\frac{1}{3} f(\phi) e^{-2(A+B)}  {A_t'}^2\big) dr \cr
& +   e^{4A(r_m)-B(r_m)}h(r_m)\bigg[3A'(r_m)\log \big\vert  \frac{h(r_m) e^{2A(r_m)}}{\alpha ^{2}}\big\vert+\frac{h'(r_m)+2A'(r_m)h(r_m)}{h(r_m)}\bigg] + \frac{10}{3} M^2 ,
\end{split}\ee 
where prime represents the derivative with respect to $r$. $\alpha$ is a free parameter that has been entered here due to the null joint term, the intersection of null sheets at $r_m$. It is in fact related to the ambiguity in normalizing the normal vectors to the null sheets, as discussed in \cite{Lehner:2016vdi}. As discussed in \cite{Carmi:2017jqz} we set $\alpha=L T$ where $L$ which is the AdS radius has already been chosen one in the paper. It should be emphasized here that $\frac{d C_A}{dt}$ reduces to eternal AdS black hole result if we initially set $Q\rightarrow 0$ and then send $r$ to zero.

Before discussing the numerical result for complexity rate we would like to study its late time behaviour. As mentioned before it can be obtained by sending $r_m$ to $r_H$. We follow the same procedure discussed in \cite{Carmi:2017jqz} and get
\be\label{MQ}
\frac{d C_A(\infty)}{dt} =  \frac{2 M^2 (Q^2 + 3 \sqrt{4 M^2+Q^4})}{Q^2 + \sqrt{4 M^2+Q^4}}\equiv M_{Q} .
\ee
We have to emphasize that for $Q\rightarrow 0$ we get the same result as \cite{Carmi:2017jqz}, $\frac{2 \hat{M}}{\pi}$, which in our notation is $\frac{6 M^2 V^{(3)}}{16 \pi^2 G_N}$.

We have presented the analytical results up to here. In the next subsection we will examine these results for different values of the field theory parameters and specifically will see how they behave near the critical point.


\subsection{Complexity=Volume Prescription}
This prescription states that the complexity is in fact the volume of the maximal codimension-one bulk time slice anchored at boundary at times $t_L$ and $t_R$. Similar to the action principle we assume $t_L=t_R$ and therefore due to our setup symmetry we expect that the complexity depends on the boundary time $t=t_R+t_L$. How to evaluate the volume of this extremal surface has been discussed in the literature such as \cite{Carmi:2017jqz} therefore we won't go through the details here and report the results in our set-up.  

The metric in Eddington-Finkelstein coordinate, $v=t+r^*(r)$, becomes
\be
ds^2 = - e^{2 A(r)} h(r) dv^2 + 2 e^{A(r)+B(r)} dv dr + e^{2 A(r)} d{\vec{x}}^2~.
\ee
Parametrizing the coordinates as $r(\lambda)$ and $v(\lambda)$, the volume of the a codimension-one surface becomes
\be
V = V_3 \int d\lambda e^{3 A(r)} \sqrt{-e^{2 A(r)} h(r) \dot{v}^2 + 2 e^{A(r)+B(r)} \dot{v} \dot{r}}~,  
\label{volume1}
\ee
where $V_3$ is the volume of three dimensional space, $\vec{x}$ and dot represents derivative with respect to $\lambda$. This relation should be extremized in order to get the volume of the maximal surface. To do this there are two observations which help us through. It is easily seen that the integral is reparametrization invariant and so we can choose $\lambda$ in a way that the integrand becomes one 
\be
\mathcal{L}(r,\dot{r},\dot{v}) = e^{4 A(r)} \sqrt{- h(r) \dot{v}^2 + 2 e^{B(r)-A(r)}  \dot{v} \dot{r}} = 1~.
\label{lag}\ee
 On the other hand the integrand does not depend on coordinate $v$ itself and only on its derivative $\dot{v}$. Therefore a conserved quantity $E$ can be defined as $\frac{- \partial \mathcal{L}}{\partial \dot{v}}$. Using \eqref{lag} we have
\be
E = e^{8 A(r)} ( h(r) \dot{v} - e^{B(r)-A(r)} )~.
\label{conserved}\ee
The above relations can be used to determine $r$ 
\be
\dot{r}^2 e^{2 B(r)+6 A(r)} = h(r) + e^{-8 A(r)} E^2~,
\label{rdot}\ee
where at $r_{min}$ we have $\dot{r}_{min}=0$ and therefore $r_{min}$ in terms of $E$ is
\be
h(r_{min}) + e^{-8 A(r_{min})} E^2 = 0~.
\label{rmin}
\ee
Note that since $r_{min}$ is smaller than horizon radius we conclude that $E< 0$. The volume is the integral over radial radius from $r_{min}$ to $r_{max}$ and therefore using \eqref{rdot} the volume relation \eqref{volume1} becomes
\be
V = 2 V_3 \int_{r_{min}}^{r_{max}} dr \frac{e^{B(r)+ 3 A(r)}}{\sqrt{h(r) + e^{-8 A(r)} E^2}}~.
\label{volume}
\ee
On the other hand one can use relation \eqref{conserved} to obtain $\dot{v}$ in terms of $E$ and $\dot{r}$ and integrate it to get
\be
t_R + r^*_\infty - r^*(r_{min}) = \int_{r_{min}}^{r_{max}} dr \bigg{[} \frac{e^{B(r)-A(r)} E}{h(r) \sqrt{E^2 + e^{8 A(r)} h(r)}} + \frac{e^{B(r)-A(r)} }{h(r)}\bigg{]}~,
\label{extra}
\ee
where due to the symmetric configuration of our set-up we have set $t|_{r_{min}}=0$. Using relation \eqref{volume} and \eqref{extra} and after some algebraic calculations the volume of the extremal or maximal surface is obtained
\be
\frac{V}{2 V_3} = \int_{r_{min}}^{r_{max}} dr e^{B(r)-A(r)} \bigg{[} \frac{\sqrt{E^2 + e^{8 A(r)} h(r)}}{h(r)} + \frac{E}{h(r)}\bigg{]} - E (t_R + r^*_\infty - r^*(r_{min}))~. 
\label{maximal}\ee
Therefore complexity in volume prescription is obtained since ${
\cal{C}}_V=\frac{V}{G_5 l}$. Having $V$ written in the above form helps us calculating the complexity rate as time derivative of ${\cal{C}}_V$. It is simple to see
\be
\frac{d {\cal{C}}_V(t)}{dt} = -\frac{V_3}{G_5 l} E = \frac{V_3}{G_5 l} e^{4 A(r_{min})} \sqrt{- h (r_{min})}~.
\ee
In order to understand how complexity varies with time it is more convinient to work with dimensionless parameters $s=\frac{r}{r_H}$ and $a \equiv \frac{dC_V}{dt}=\frac{G_5 l}{V_3} \frac{d{\cal{C}}_V}{dt}$. In terms of $s$ we have
\be
a = s_{min}^{\frac{5}{3}}(s_{min}^2+\frac{Q^2}{r_H^2})^{\frac{1}{6}}\sqrt{(1-s_{min}^2)(s_{min}^2+1+\frac{Q^2}{r_H^2})}~.
\ee %
Therefore it is straightforward to write the equation \eqref{extra} in terms of these dimensionless variables. The final result is 
\be\label{cvrate}\begin{split}%
r_H t  = 2 a \int_{s_{min}}^{\infty} ds & \frac{s\sqrt{s^2+\frac{Q^2}{r_H^2}}}{(1-s^2)(s^2+1+\frac{Q^2}{r_H^2})} \times \cr
&\bigg[s_{min}^{\frac{10}{3}}(s_{min}^2+\frac{Q^2}{r_H^2})^{\frac{1}{3}}(1-s_{min}^2)(s_{min}^2+1+\frac{Q^2}{r_H^2})
 -s^{\frac{10}{3}}(s^2+\frac{Q^2}{r_H^2})^{\frac{1}{3}}(1-s^2)(s^2+1+\frac{Q^2}{r_H^2})\bigg]^{-\frac{1}{2}},
\end{split}\ee%
where $t_R = t/2$. 

To study the late time behaviour of complexity rate we need to obtain the value of $r_{min}$ at $t \rightarrow \infty$. From equation \eqref{rmin} we know that $r_{min}$ is the root of $e^{4 A(r_{min})} \sqrt{- h (r_{min})}=\pm E$. There are two roots for $r_{min}$ where as $t \rightarrow \infty$ they coincide and therefore the late time value of $r_{min}$ is the root of both equations \eqref{rmin} and $\frac{d}{dr} (e^{4 A(r)} \sqrt{- h (r)})=0$.

\section{Numerical Results}
\subsection{Action Prescription}
In this section we will study the behaviour of the previously obtained results for different values of the parameters relevant to field theory. We are, in particular, interested to see how the field theory quantities behave as we move towards the critical point in the field theory phase space, $\frac{\mu}{T}^*$. 
\begin{figure}
\includegraphics[width=90mm]{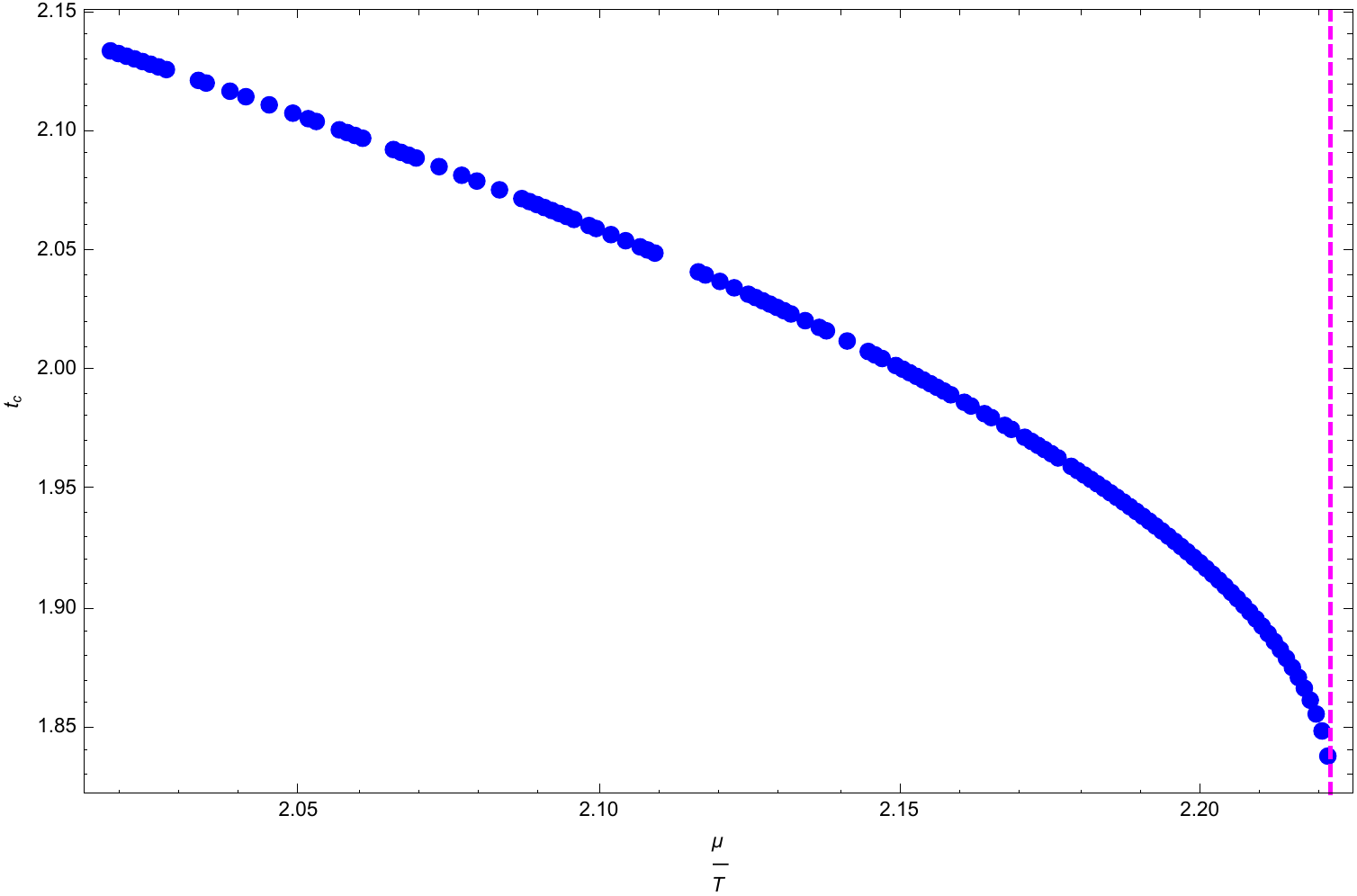}
\caption{The behaviour of the critical time $t_c$ with respect to $\frac{\mu}{T}$. The magenta dashed line shows the critical value $\m^*$.} \label{tc}
\end{figure} 
\begin{figure}
\centering
\includegraphics[width=83mm]{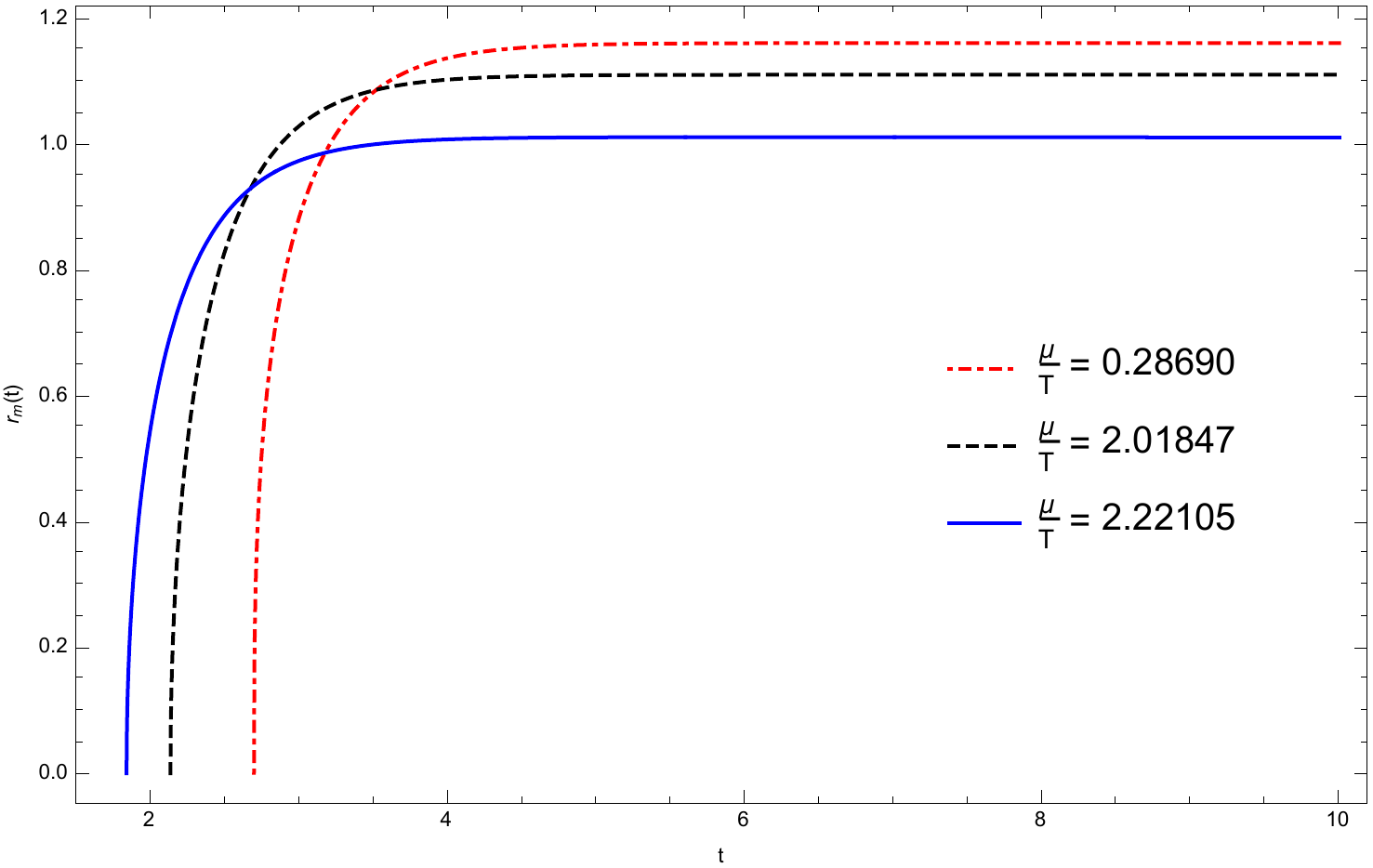}
\includegraphics[width=94mm,trim = 0 0.41cm 0 0]{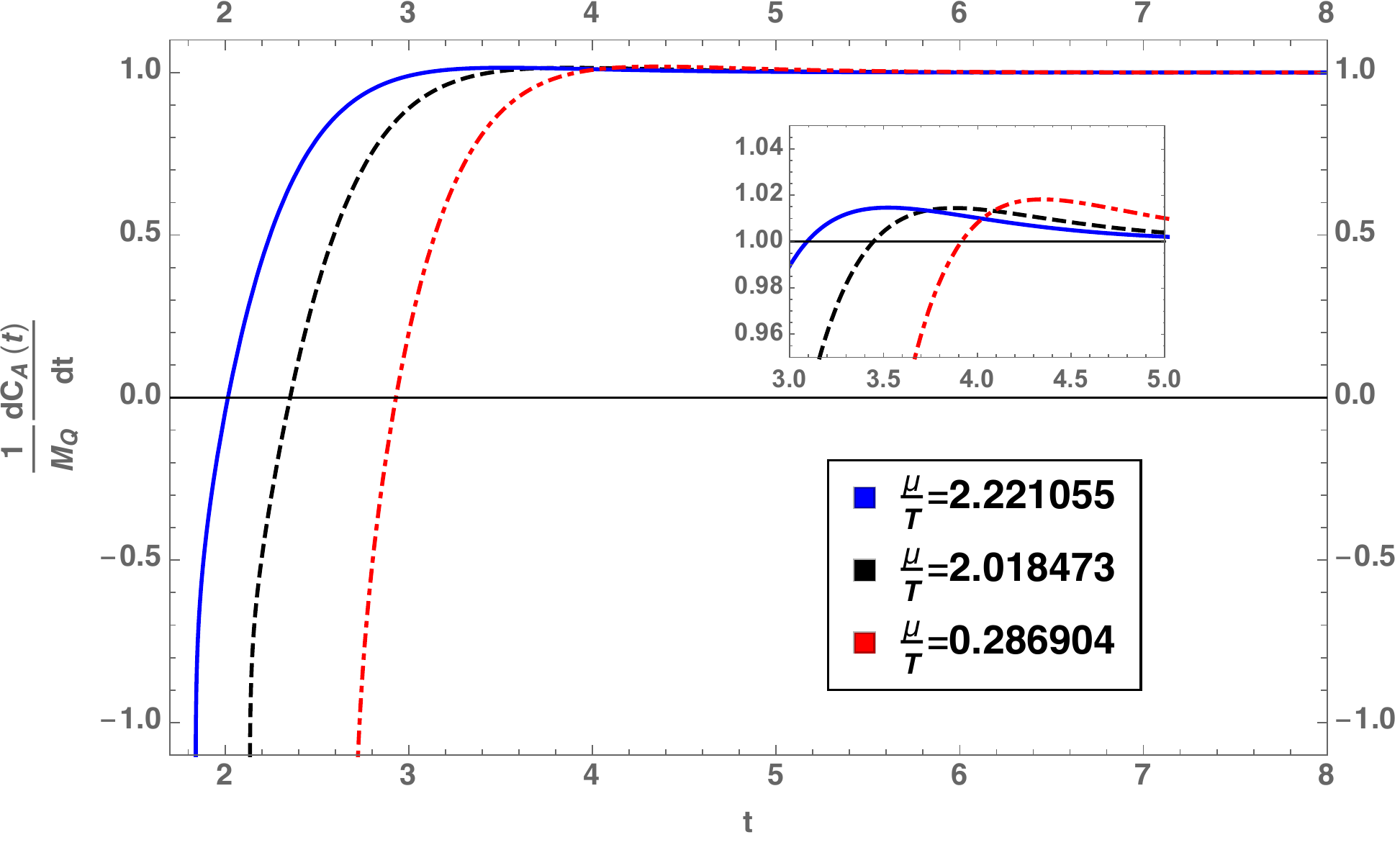}
\caption{Left: The behaviour of $r_m$ with respect to time for different values of $\m$. Right: The complexity rate for different values of $\m$.} \label{complexity}
\end{figure}  
Let us first consider the critical time, $t_c$. Knowing how $r^*$ looks like in terms of the coordinate $r$, \eqref{rs}, we can obtain how the critical time, \eqref{critical t}, evolves as we move nearer to the critical point. This has been plotted in figure \ref{tc}.  Please note that we have set $T=0.37$ for all the results in this section. A quick conclusion from this figure is that as we move towards the critical point in the parameter $\frac{\mu}{T}$ the critical time decreases and reaches a minimum finite value with an infinite slope. The magenta dashed line shows $\frac{\mu}{T}^*$. Reminding ourselves that the critical time is the time at which the complexity rate becomes non-zero, we can conclude that the complexity of a thermofield double state produced in field theory starts changing with time sooner as $\m$ moves closer to $\frac{\mu}{T}^*$. Equivalently the complexity of formation of this state in the boundary theory remains constant for longer time as $\m$ decreases. In other words the critical time, $t_c$, decreases in the presence of chemical potential and it gets its minimum value approaching the critical point.  
\begin{figure}
\centering
\includegraphics[width=90mm]{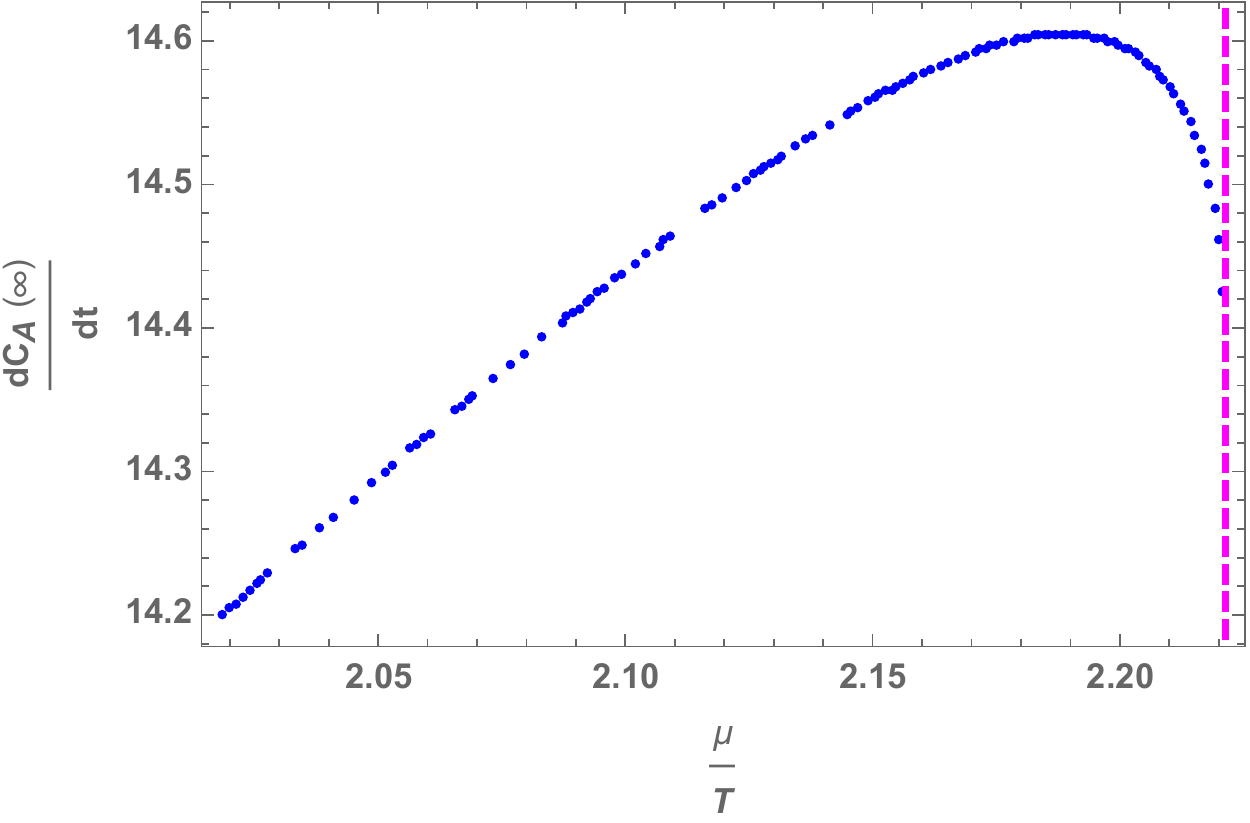}
\includegraphics[width=87mm,trim = 0 -1.6cm 0 0]{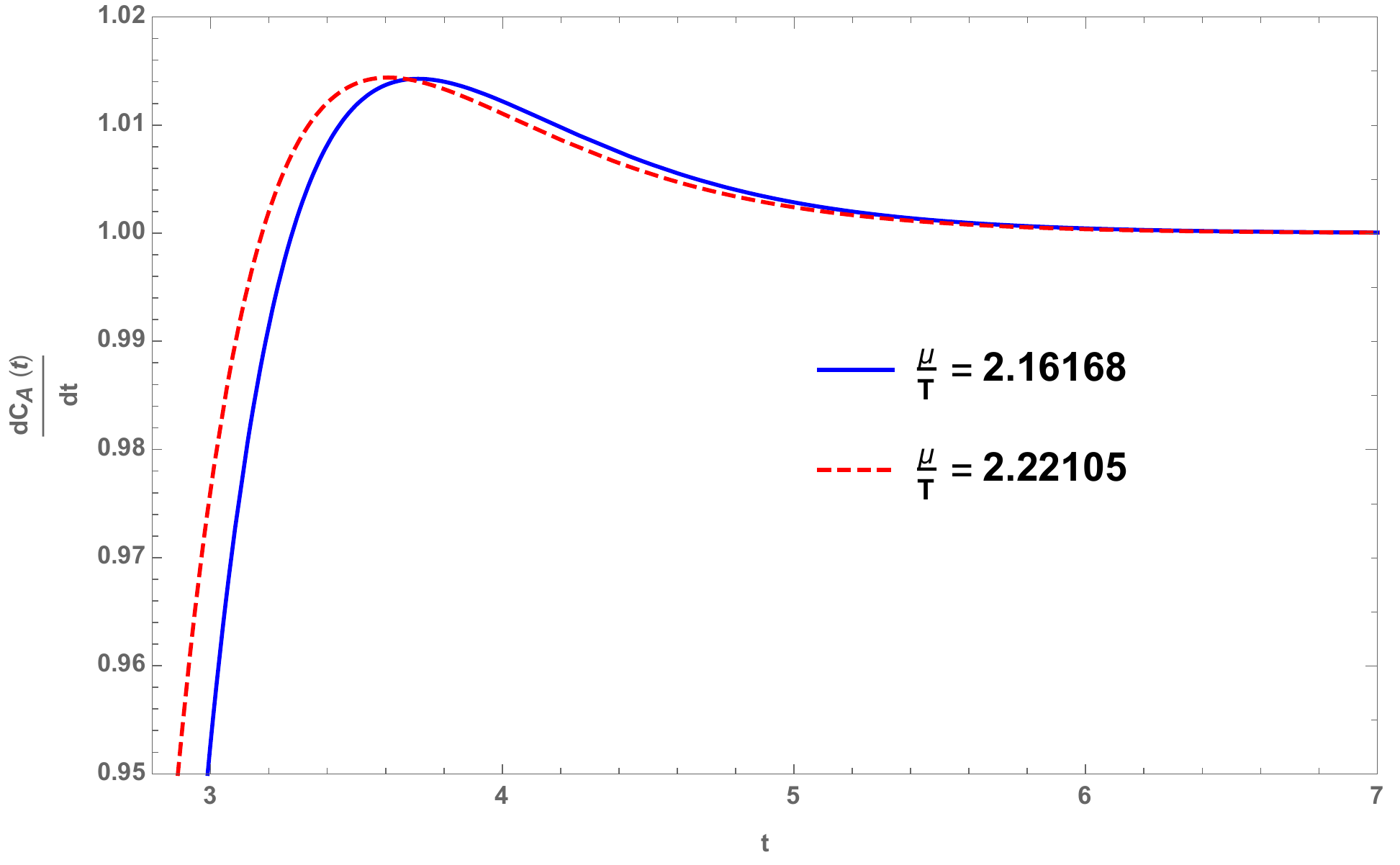}
\caption{Left: The final value of complexity rate in terms of $\m$. The magenta dashed line presents $\m^*$. Right: Complexity rate for two different $\m$ with the same late time value} \label{finalc}
\end{figure} 
\begin{figure}
\centering
\includegraphics[width=85mm]{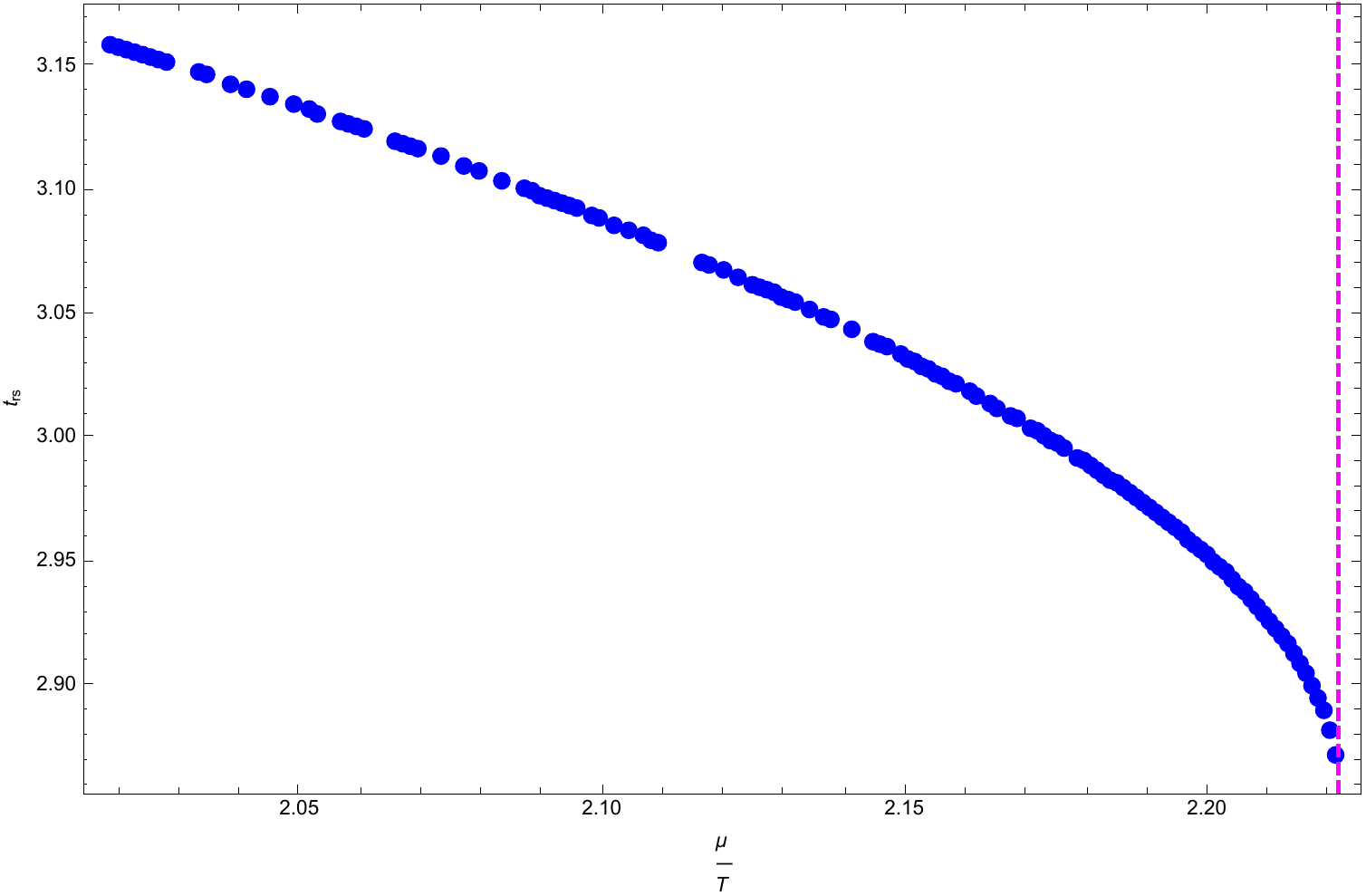}
\includegraphics[width=91mm,trim = 0 0.41cm 0 0]{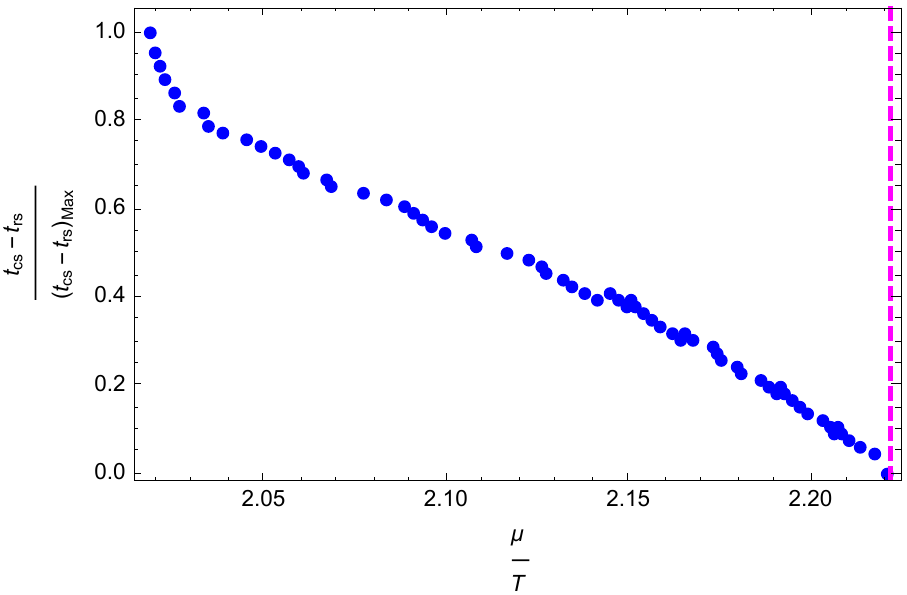}
\caption{Left: The behaviour of saturation time $t_{rs}$ with respect to $\frac{\mu}{T}$. The magenta dashed line shows the critical value $\m^*$. Right: The numerical difference between the saturation times, $t_{cs}-t_{rs}$ is plotted with respect to $\frac{\mu}{T}$. It can be almost fitted with a linear function.} \label{trs}
\end{figure} 

In figure \ref{complexity}, left panel, we show the result of solving the differential equation \eqref{eq1} numerically by setting the initial condition for $r_m(t_c)$ to be very small or close to singularity such as $0.001$ for different values of $\frac{\mu}{T}$. It is obvious that $r_m$ starts from an initial value near the past singularity, at $t=t_c$, and increases until it saturates to a final value which is equal to $r_H$ as it is expected from equation \eqref{eq1}. It can be clearly seen in the figure that the final value of $r_m$ decreases for larger $\m$ and therefore we expect that its minimum value occurs at $\m^*$.

The next quantity that its dependence on $\m$ can be studied is the complexity rate $\frac{dC}{dt}$. It is plotted in figure \ref{complexity}, right panel. This figure shows that after $t>t_c$ the complexity rate increases up to its maximum value in a short time interval and then relaxes to a constant quantity at late times. As it can be seen in the figure complexity rate reaches the final value from above and similarly to the results reported in the literature the Lloyd's bound, \cite{lloyd}, is violated \cite{Carmi:2017jqz}. The Lloyd's bound refers to an upper bound on the rate of complexification of a system with mass $M$; i.e. $\frac{dC}{dt} \leq \frac{2 M}{\pi}$. The violation of Lloyd's bound in holographic studies of complexity has been discussed in two manners, one where the bound is violated at early time and then the complexity rate approaches it at late time such as in \cite{Carmi:2017jqz} and references therein and the other one where the bound is violated both in early time and late time such as in \cite{Mahapatra:2018gig}. In our case where in holographic background the charge as well as the mass is not zero we name the final value that the complexity rate reaches the Lloyd's bound and will observe that this bound is violated at early time and reaches the bound from above at late time. We have to report that for the background we have considered this violation is very small. The constant late-time value of complexity rate in our case depends on $Q$ and $M$ and varies as $\m$ changes based on $M_Q$ in \eqref{MQ}. We would also like to emphasize that figure \ref{complexity}, right panel, includes a negative peak at early times. Although we have not produced this behaviour analytically, it has been previously observed in AdS-Schwarschild black hole and other asymptotically AdS black holes and reported in several papers such as \cite{Carmi:2017jqz} and \cite{Mahapatra:2018gig}.

We have examined the late time behaviour of complexity rate more closely in two different ways. On the one hand we can use the analytic result $M_Q$ in \eqref{MQ} which is obtained from late time expansion of $\frac{dC}{dt}$ in \eqref{rate}. On the other hand we can use the numerical result and plot the final values that complexity rate reaches at late time for different $\m$. Both of these give the same result as shown in figure \ref{finalc}, left panel. The intriguing observation is that the growth of complexity rate at late times peaks before the critical $\m$, magenta dashed line in plot, at  about $\m\approx 2.18$ and then starts decreasing as we get very much closer to the critical point. Therefore two different values of $\m$ correspond to one specific value of complexity rate \footnote{We believe that $dC_A(\infty)/dt$ becomes tangent to magenta dashed line for the values of $\m$ close to $\m^*$ in figure \ref{finalc}.}. In other words there are two different states in field theory characterized by $\mu$ and $T$ with the same late-time values of complexity rate, for example figure \ref{finalc}, right panel. This somehow resembles {\it{memory loss}} where the late-time complexity rate alone is not enough to fully specify the state. Such behaviour in late-time complexity rate happens in contrast to the $\m$ dependence of the late-time values of $r_m$ where it continuously decreases moving towards the critical point and reaches its minimum value there. It should also be mentioned that complexity rate in the field theory dual to our background is different from the one dual to usual asymptotically AdS charged black hole backgrounds obtained from Einstein-Maxwell theory in the sense that the presence of non-zero chemical potential does not modify the early time behaviour as we still can see the negative divergent complexity rate there \cite{Carmi:2017jqz}. 

As mentioned before the background we have been studying is dual to field theory at finite temperature and chemical potential and there exists a critical point in the phase diagram of the theory. Due to the presence of critical point the field theory thermodynamic quantities diverge as one moves towards it and one can read the critical exponent from the way these parameters diverge at the critical point. Using the Kubo commutator of conserved currents one can read dynamical critical exponent and classify theories accordingly \cite{DeWolfe:2011ts}. The dynamical critical exponent has also been obtained in the field theory dual to this background in \cite{Finazzo:2016psx} and \cite{Ebrahim:2017gvk} using the quasi-normal modes and equilibration time in field theory, respectively. We can now ask this question whether information on complexity provides us some equilibration times in order to obtain the critical exponent. Putting it in other words we intend to demonstrate how using complexity one can detect some important physical information in field theory. Therefore we define two saturation times as follows:
\begin{itemize} 
\item The first quantity we define is the saturation time for $r_m$ which is represented as $t_{rs}$ here and defined as the time after which the relation $|r_m(\infty)-r_m(t_{rs})|< 0.05~r_m(\infty)$ is always satisfied. It has been plotted in figure \ref{trs}, left panel, with respect to $\m$. The behaviour is quite similar to figure \ref{tc}; as we move towards the critical point, $r_m$ saturates and reaches its final value in smaller timescales. Comparing these two figures, one can easily conclude that the smaller $t_c$, the smaller $t_{rs}$. 

\item The other saturation time is defined as the saturation time for the complexity rate, called $t_{cs}$.  We have defined it numerically similarly to $t_{rs}$ as a time at which the quantity $|\frac{dC(\infty)}{dt}-\frac{dC(t_s)}{dt}|<0.05 \frac{dC(\infty)}{dt}$ and stays blew this limit afterwards. The result for saturation time of complexity rate is similar to $t_{rs}$ plotted in figure \ref{trs}, left, so we don't include it here. Moving towards the critical point the saturation time decreases with a seemingly infinite slope. 

It is interesting to examine if the relaxation of $r_m$ and complexity rate to their late-time values happen at the same time. The result has been plotted in figure \ref{trs}, right panel, where the numerical difference, $t_{cs}-t_{rs}$ has been scaled with its maximum value in the plot. The first observation is that this quantity decreases by raising $\m$ almost linearly. It's also illuminating to see that as one moves towards the critical point, magenta dashed line in plot, this difference becomes less and less significant until it gets infinitesimally small value or becomes almost zero. {\it{Therefore $r_m$ and $\frac{dC}{dt}$ saturate at almost the same time very close to the critical point, $\m^*$}}. 
\end{itemize}
\begin{figure}
\centering
\includegraphics[width=85mm,trim = 0 0 0 0]{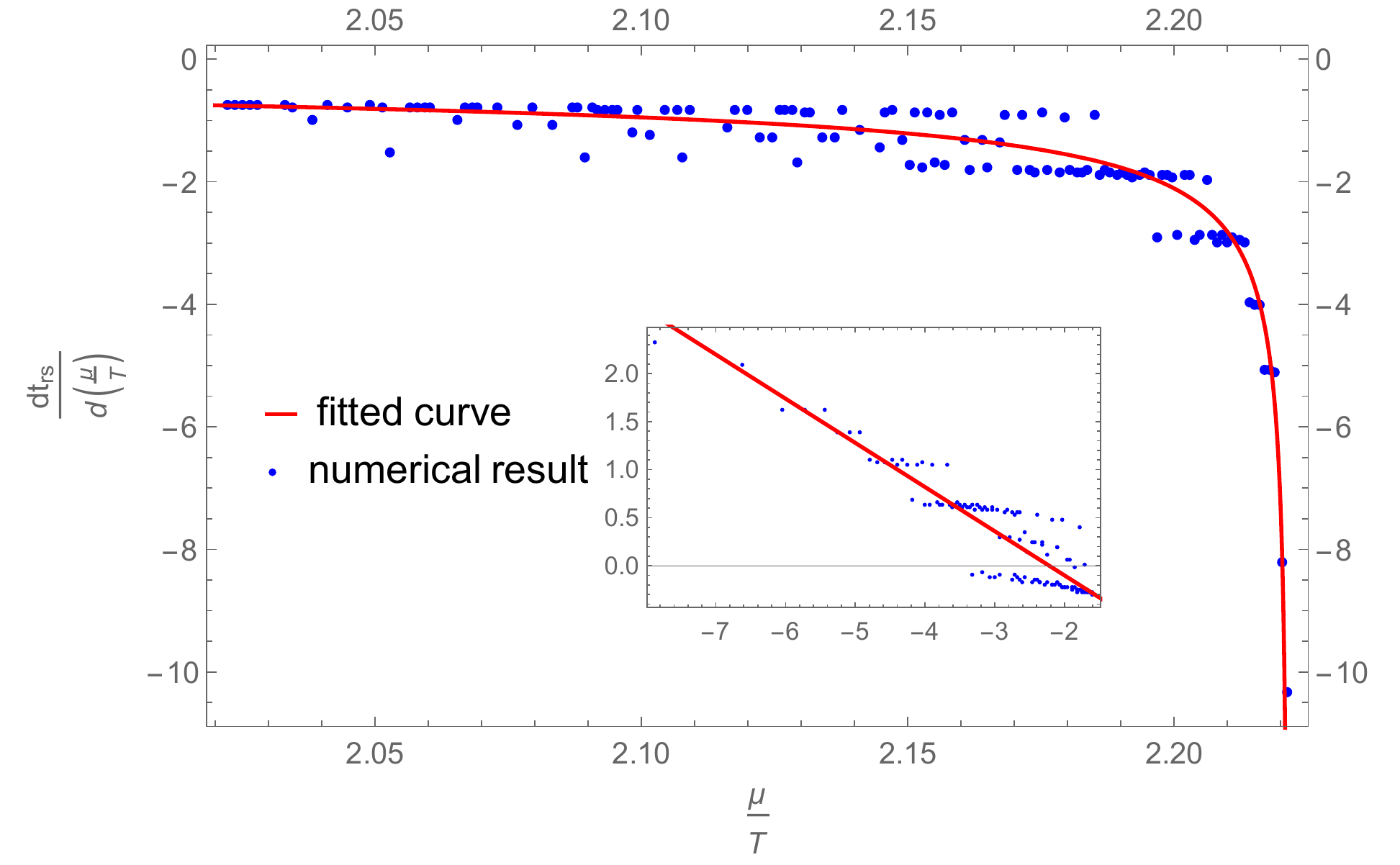}
\includegraphics[width=85mm]{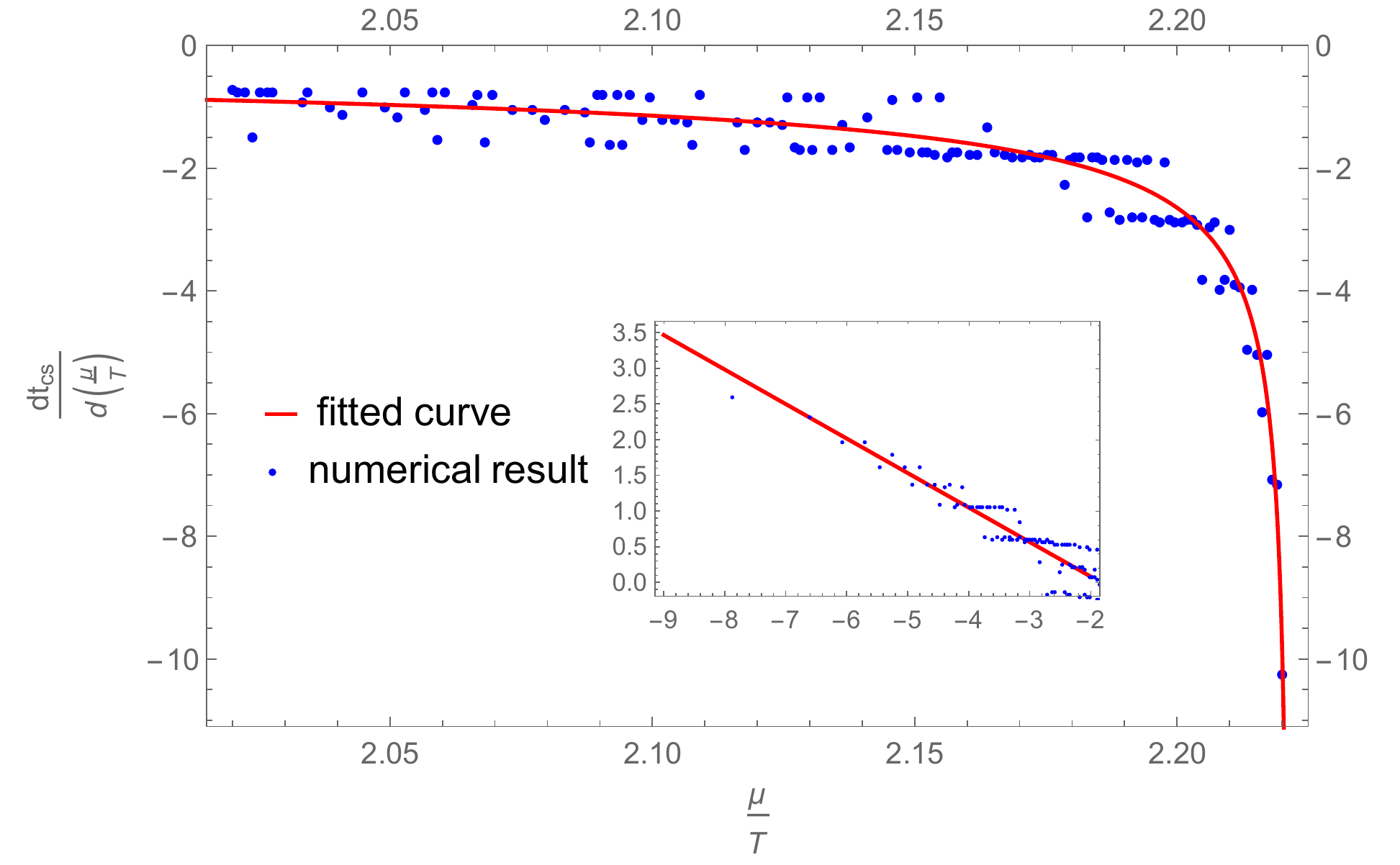}
\caption{Left: The slope of the saturation time for $r_m$, $t_{rs}$, with respect to $\frac{\mu}{T}$. The fitted curve with the numerical result is $-0.4 (\frac{\pi}{\sqrt{2}}-\frac{\mu}{T})^{-0.46}$ with rms=0.07. Right: The slope of the saturation time for $\frac{dC}{dt}$, $t_{cs}$, with respect to $\frac{\mu}{T}$. The fitted curve with the numerical result is $-0.4 (\frac{\pi}{\sqrt{2}}-\frac{\mu}{T})^{-0.48}$ with rms=0.22. The small plot is the logarithm of the data results and the linear function fitted with them. } \label{exp}
\end{figure} 

As it was mentioned in the previous paragraph, looking closely at figure \ref{trs}, left, and the identical one for $t_{cs}$ which we have not plotted here, the behaviour of $t_{rs}$ and $t_{cs}$ near the critical point are the same. They both decrease moving towards the critical point and reach a finite value there but the slope of the change in time with respect to $\m$ seems to go to infinity. It's intriguing to check how the slope changes as a function of $(\m^*-\frac{\mu}{T})$. One can assume that if the slope function fits with $(\m^*-\frac{\mu}{T})^{-\theta}$ the value of $\theta$ gives the dynamical critical exponent in the field theory dual to this background. The dynamical critical exponent obtained using the Kubo commutator for conserved currents in the field theory dual is $0.5$ \cite{DeWolfe:2011ts}.  In fact this has been checked and confirmed in the papers \cite{Finazzo:2016psx} and \cite{Ebrahim:2017gvk} where the authors have studied the quasi normal modes and  equilibration time behaviour near the critical point, respectively. 
\begin{figure}
\centering
\includegraphics[width=85mm]{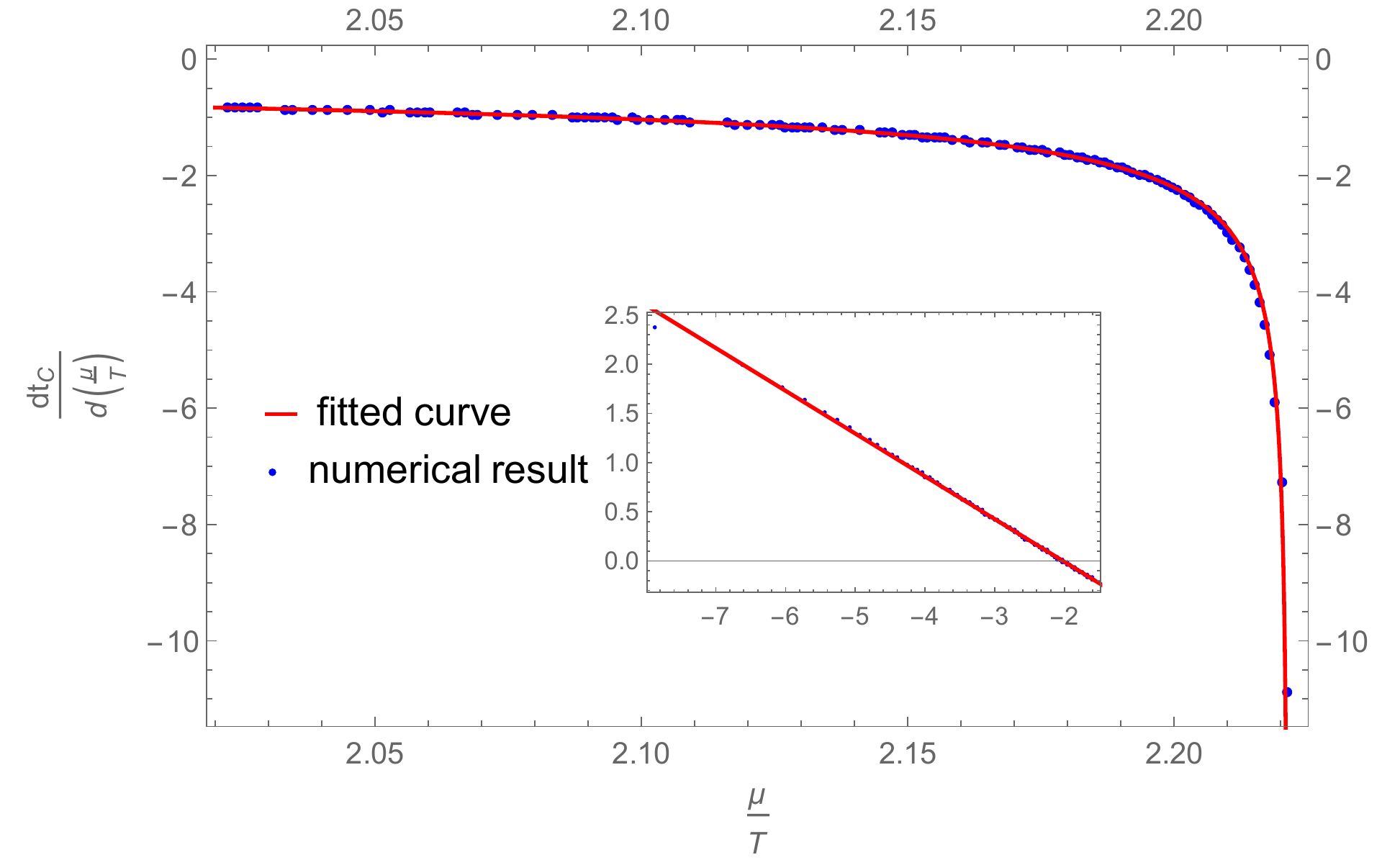}
\caption{The slope of the critical time $t_c$ with respect to $\frac{\mu}{T}$. The blue dots are the numerical result and the red curve is the fitted function as $-0.4 (\frac{\pi}{\sqrt{2}}-\frac{\mu}{T})^{-0.44}$ with rms=0.02. The small plot is the logarithm of the data results and the linear function fitted with them.} \label{exptc}
\end{figure} 

The slope near the critical point is evaluated using 
\be
\frac{dJ}{d \frac{\mu}{T}} (i) = \frac{J (i+1) - J (i)}{\frac{\mu}{T} (i+1) - \frac{\mu}{T} (i)},
\label{slope}
\ee
where $J$ is $t_{rs}$ or $t_{cs}$. The results have been plotted in figure \ref{exp}. The blue points show the numerical result and the red curve is the fitted function $(\m^*-\frac{\mu}{T})^{-\theta}$. The results for $\theta$ are $0.46$ and $0.48$ for plots left and right in figure \ref{exp}, respectively. This is a very interesting result as, although we have calculated $\theta$ from two basically different quantities but their values are very close and also very close to $0.5$, dynamical critical exponent known for this background \cite{DeWolfe:2011ts}. One of these quantities is the complexity rate saturation time and the other one is the saturation time at which $r_m$ relaxes to its final value which is $r_H$, the horizon radius of the background. 

We would like to give more details on our fitted functions used to read the critical exponent. In order to obtain a better fit for our result we fit a linear function with the logarithm of both sides of the equation $\frac{dJ}{d \frac{\mu}{T}} \propto (\m^*-\frac{\mu}{T})^{-\theta}$ in our data points. In order to report how well our fitted $\theta$ is we also calculate root mean square (rms) and have reported it in the caption of each plot. For a linear function $y=a x+b$ the rms is defined as 
\be
rms = \sqrt{\frac{1}{N} \Sigma (y_{fit}-y_{data})^2}~,
\ee
where $y_{fit}$ is the value of fitted function $y$ evaluated at data points $x$ and $y_{data}$ is the corresponding value read from data and $N$ is the number of data points. The smaller rms is the better the function fits with data.   

It is interesting to see if the other quantities calculated in this background behave similarly near the critical point. One such quantity is $t_c$ or critical time plotted in figure \ref{tc}. We follow the same procedure done for $t_{rs}$ and $t_{cs}$ to obtain the critical exponent $(\theta=0.44)$ where the result is plotted in figure \ref{exptc}. Although $t_c$ by its nature is not defined as a saturation time, interestingly, it produces similar result for critical exponent to $t_{rs}$ and $t_{cs}$ with a good approximation. We don't a priori know the reason why critical time can produce similar critical exponent.       

Before closing this section we would like to comment on how our results relate to other papers on this subject. In paper \cite{Zhang:2017nth} the behaviour of complexity near critical point in the QCD model \cite{Gubser:2008ny} has been discussed. This model has a critical point at temperature $T_c$ and differs from the model we discuss here where we have nonzero chemical potential as well as temperature. The author in \cite{Zhang:2017nth} observes a jump in $\frac{dC(\infty)}{dt}$ at critical temperature while we don't see such behaviour in our case as has been shown in figure \ref{finalc}, left. This difference might be due to the presence of nonzero chemical potential. In addition to the complexity rate we obtain the dynamical critical exponent of the model using the data near the critical point. 

 We should also point out an important difference between our result here and the one obtained in the literature for AdS-RN black hole solutions. As it has been discussed in \cite{Carmi:2017jqz} due to the presence of two horizons in such solutions the complexity rate is nonzero all the time for large enough $\m$. In other words for large enough charge there is no $t_c$ before which the time derivative of complexity vanishes. But in our case the complexity rate is zero before critical time, $t_c$, as discussed in the previous paragraph. 

\subsection{Volume Prescription} 
\begin{figure}
\includegraphics[width=88mm, trim = 0 -1.2cm 0 0]{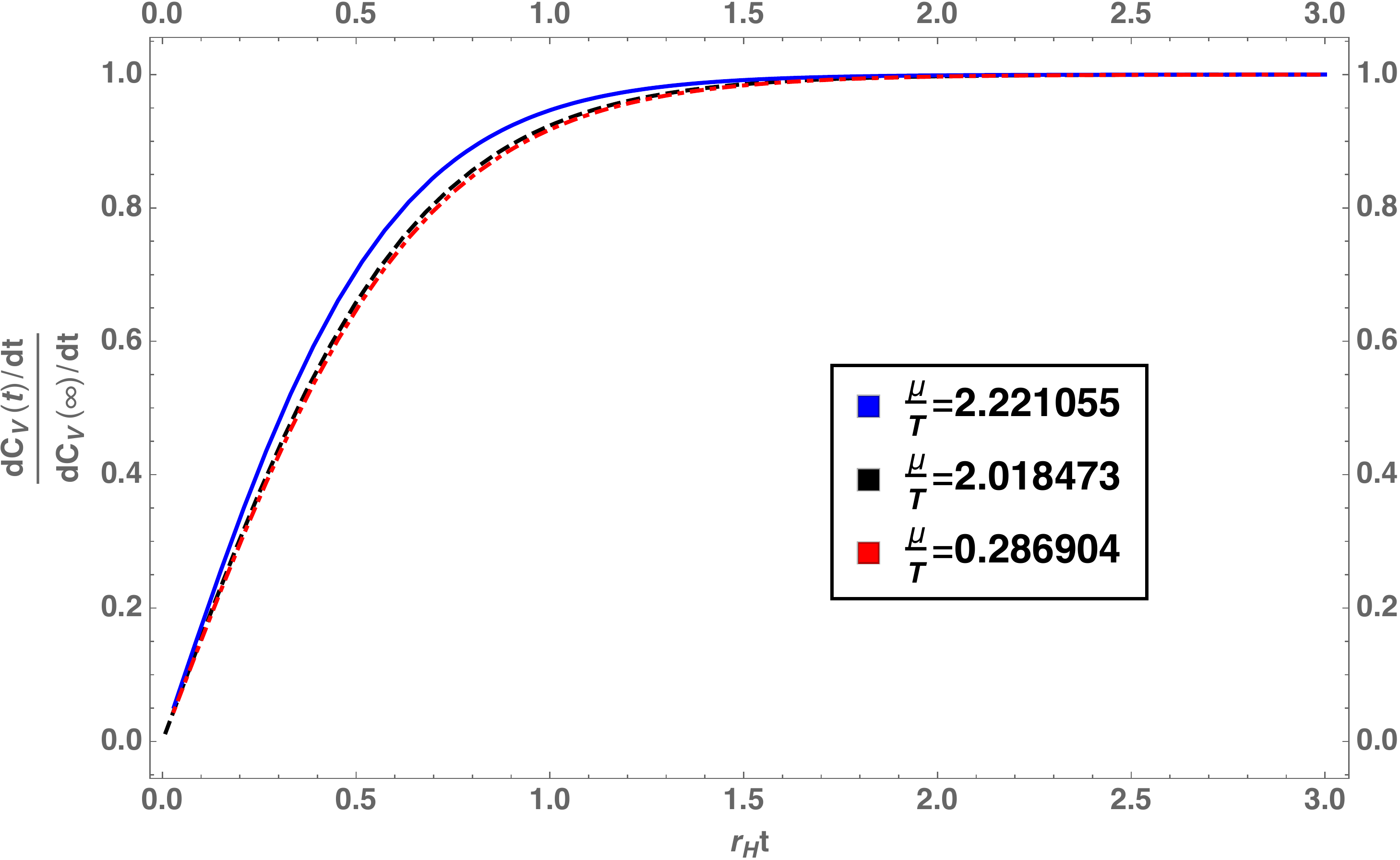}
\includegraphics[width=87mm]{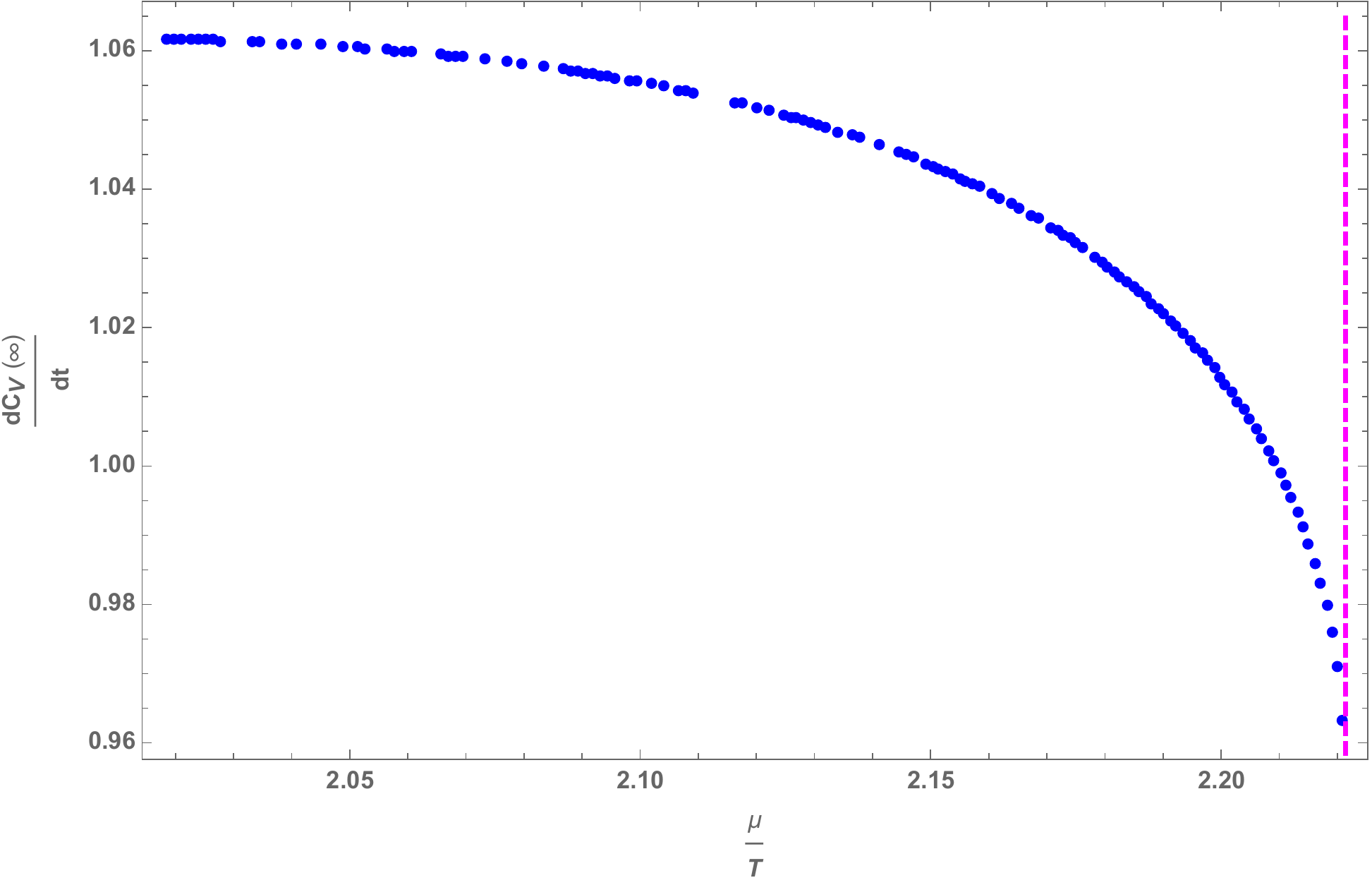}
\caption{Left: The complexity rate in volume prescription for different values of $\m$. Right: The late time behaviour of complexity rate in volume prescription with respect to $\m$. The magenta dashed line in $\m=\m ^*$.} \label{latecv}
\end{figure}

We are interested to see what the differences between the numerical results in action prescription and volume prescription are. The rate of complexity in volume prescription for different values of $\m$ has been plotted in figure \ref{latecv}, left panel, using \eqref{cvrate}. We have plotted the rescaled complexity rate with its late time value with respect to the dimensionless quantity $r_H t/L$ where we have set $L=1$. The figure is independent of the ratio $r_H/L$. It can be compared with figure \ref{complexity}, right panel where we have plotted complexity rate in action prescription for various values of $\m$. It is clear that in contrast to complexity rate in action prescription, the complexity rate in volume prescription is less sensitive to $\m$ and being very close to the critical point does not make much difference. 

We could also check whether we can obtain a similar critical exponent using the volume prescription to the action prescription. It is clear from figure \ref{latecv}, left panel, that there will be very small difference between saturation time of complexity rate for different values of $\m$. Therefore our numerical evaluation indicates that we will not be able to obtain critical exponent from volume prescription.  

The other main difference between these two prescriptions is their late time behaviour. We plot the late time behaviour of complexity rate in volume prescription in terms of $\m$ in figure \ref{latecv}, right panel, and it can be compared with late time complexity rate in action prescription, figure \ref{finalc} left panel. It can be easily seen that late time complexity rate in volume prescription continuously decreases as we move towards the critical point while in action prescription it increases first and after a peak it starts decreasing.

\section*{Acknowledgement}
We would like to thank M. M. Sheikh-Jabbari, M. R. Mohammadi Mozafar, R. Fareghbal and M. Lezgi for fruitful discussions and comments. M. Asadi is a post-doc fellow of Boniad Melli Nokhbegan (BMN) under Allameh Tabatabayie grant c/o M.M. Sheikh-Jabbari.

\appendix
\section{Some details on complexity analytic calculations}\label{appA}
As it was discussed in \ref{csection} the complexity in its action prescription can be evaluated by calculating the action \eqref{actionf} on WdW patch in figure \ref{WdWp}. The time-dependent part of the complexity is the one that is important for us since it results in complexity rate. We will give some details on how to calculate it in the following:
\begin{itemize}
\item{\bf Bulk Contribution}\\
It was mentioned previously that the first line in the gravitational action \eqref{actionf} gives the bulk contribution. The curvature $R$ in the background we consider here is $- \frac{d(d+1)}{L^2}=- 20$ since we have set $L=1$. Therefore we have 
\bea
I_{\text{bulk}} &=&\frac{ V_{3}}{16\pi G_5}\int d\tau dr \sqrt{-g}  \bigg(R-\frac{f(\phi)}{4} F_{\mu\nu} F^{\mu\nu}-\h\partial_\mu \phi \partial^\mu \phi-V(\phi)\bigg)\cr  
&=& \frac{ V_{3}}{16\pi G_5}\int d\tau dr \bigg(-16 r^3- \frac{8r^3}{3}(\frac{r^2+Q^2}{r^2})+\frac{4M^2Q^2}{3 r^3}(\frac{r^2+Q^2}{r^2})^{-2}\bigg)\\
&=& \frac{ V_{3}}{16\pi G_5}\int d\tau dr \, {\cal{I}}(r)
\eea
where $V_{3}$ is the volume of the spatial geometry. Due to the symmetry between left and right sides of WdW patch we can evaluate this integral on the right side and then multiply it by two. We will repeat this in the other calculations too. Different parts of this patch, as denoted by I, II and III in figure \ref{WdWp}, produce
\bea
I_{\text{bulk}}^{I}& = & \frac{ 2V_{3}}{16\pi G_5} \int_{0}^{r_h} dr\, {\cal{I}}(r) \,\int_0^{t_R+r^*_\infty-r^*(r)} d\tau 
=\frac{ V_{3}}{8\pi G_5} \int_{0}^{r_h} dr\, {\cal{I}}(r) \, (t_R+r^*_\infty-r^*(r)) ,\\
I_{\text{bulk}}^{II} & = &\frac{ 2V_{3}}{16\pi G_5} \int_{r_h}^{r_{\text{max}}}  dr\, {\cal{I}}(r) \, \int_{t_R-r^*_\infty+r^*(r)}^{t_R+r^*_\infty-r^*(r)} d\tau  
=\frac{ V_{3}}{4\pi G_5} \int_{r_h}^{r_{max}} dr\,  {\cal{I}}(r) \,(r^*_\infty-r^*(r)) ,\\
I_{\text{bulk}}^{III} & = & \frac{ 2V_{3}}{16\pi G_5} \int_{r_m}^{r_h} dr \,  {\cal{I}}(r) \,\int^0_{t_R-r^*_\infty+r^*(r)} d\tau  
=\frac{ V_{3}}{8\pi G_5} \int_{r_m}^{r_h} dr\,  {\cal{I}}(r) \,(-t_R+r^*_\infty-r^*(r)) . 
\eea
As it can be seen the contribution of the part II in the action is time independent. Since we have assumed $t_R=t_L=\frac{t}{2}$ and using the left-right symmetry the total contribution of the bulk action becomes
\be
\frac{d I_{\text{bulk}}}{dt}=\frac{ V_{3}}{16\pi G_5} \int_{0}^{r_m} {\cal{I}}(r)\, dr ,
\ee
where $I_{\text{bulk}} = I_{\text{bulk}}^I + I_{\text{bulk}}^{II} + I_{\text{bulk}}^{III}$.
\item{\bf GHY Surface Term}\\
As mentioned in section \ref{csection} GHY terms are only defined on time-like or space-like surfaces. Since we are only interested in time-dependent contributions to the action the only non-zero contributions come from future singularity and the uv cut-off in asymptotic boundaries. In order to calculate the GHY part of the action we need to have the trace of the extrinsic curvature on these two $r$ constant surfaces. It is defined as $K=h^{ab} K_{ab} = \nabla_{a} n^a$ where $h_{ab}$ is the induced metric and $n_a$ is the normal vectors for these surfaces. Note that we use $a$ and $b$ letters for the coordinates on the time-like or space-like surfaces. In general for any $r=constant$ surface which is defined as $\Phi(r)=r- constant$ we have
\be
n_r=\frac{\varepsilon}{\sqrt{|g^{rr} \partial_r \Phi \partial_r \Phi|}}\bigg{|}_{r=constant} ,
\ee
where $\varepsilon=+1, -1$ for time-like and space-like surfaces, respectively. Therefore the final result for GHY action becomes 
\bea
I^{\text{future}}_{\text{surf}}&=&\frac{V_{3}}{8\pi G_5} h(r) e^{4 A(r)-B(r)} \big(8 A'(r)+\frac{h'(r)}{h(r)}\big) \int_{0}^{t_R+r^*_\infty-r^*(r)} d\tau \bigg{|}_{r=\epsilon_0},\cr
I^{\text{uv}}_{\text{surf}}&=&\frac{V_{3}}{8\pi G_5}h(r) e^{4 A(r)-B(r)} \big(8 A'(r)+\frac{h'(r)}{h(r)}\big) \int_{t_R-r^*_\infty+r^*(r)}^{t_R+r^*_\infty-r^*(r)} d\tau \bigg{|}_{r=r_{\text{max}}} ,
\eea
and we define $I_{\text{surf}}=I^{\text{future}}_{\text{surf}}+I^{\text{uv}}_{\text{surf}}$.
\item{\bf Null Joint Contribution}\\
Normal vectors to the past null boundaries are
\be
(K_L^{(\alpha)})_\mu =(-\alpha , \alpha \frac{e ^{B-A}}{h}), \qquad (K_R^{(\alpha)})_\mu =(\alpha , \alpha \frac{e ^{B-A}}{h}),
\ee
where $\alpha$ is the normalization constant which should be fixed (Discussion on ambiguities regarding the null segments can be found in \cite{Lehner:2016vdi}.). The joint action is defined as
\be
I_{\text{joint}}=\frac{1}{8\pi G_5} \int d^3x \sqrt{\sigma}\,\log \vert \frac{1}{2}K_L.K_R\vert .
\ee
Therefore we have 
\begin{align}
\begin{split}
I_{\text{joint}}&=-\frac{V_{3}}{8\pi G_5}\, \left[ e^{3A(r)}\log\vert  \frac{h(r)\,e^{2A(r)}}{\alpha ^{2}}\vert\right] \bigg{|}_{r=r_m} ,\\
\frac{d I_{\text{joint}}}{dt}&=\frac{V_{3}}{16\pi G_5}\left\lbrace e^{4A(r)-B(r)}h(r)\,\big(3A'(r)\log \vert  \frac{h(r)\,e^{2A(r)}}{\alpha ^{2}}\vert+\frac{h'(r)+2A'(r)h(r)}{h(r)}\big)\right\rbrace \bigg{|}_{r=r_m} .
\end{split}
\end{align}

\item{\bf Total Action} \\
The rate of growth of holographic complexity is given by
\be
\frac{d {\cal{C}}_A}{dt}=\frac{1}{\pi}\frac{d}{dt}(I_{\text{bulk}}+I_{\text{joint}}+I_{\text{surf}}).
\ee

\end{itemize}

\end{document}